\documentclass[a4paper,prl,superscriptaddress]{revtex4}

\usepackage{times,ulem,color,bm}
\usepackage{amsfonts}
\usepackage{epsf,subfigure}
\usepackage{amsmath,graphicx}
\usepackage[utf8]{inputenc}
\usepackage[T1]{fontenc}
\usepackage{times,ulem,color,bm}
\usepackage[squaren,Gray]{SIunits}

\newcommand{\cora}[1]{\textcolor{black}{#1}}
\newcommand{\cor}[1]{\textcolor{black}{#1}}

\usepackage[pdfstartview=FitV,bookmarks=true,
            linktocpage=true,colorlinks=true,
            urlcolor=blue,linkcolor=blue,
            citecolor=blue]{hyperref}

\begin{document}

%
%


\title{Matrix Approach of Seismic Imaging: Application to the Erebus Volcano, Antarctica}

%
%



\author{Thibaud Blondel}
\affiliation{ESPCI Paris, PSL Research University, CNRS, Univ Paris Diderot, Sorbonne Paris Cit\'{e},\\ Institut Langevin, UMR 7587, 1 rue Jussieu, F-75005 Paris, France}

\author{Julien Chaput}
\affiliation{Department of Geological Sciences, University of Texas El Paso, El Paso, TX, USA}
\author{Arnaud Derode}
\affiliation{ESPCI Paris, PSL Research University, CNRS, Univ Paris Diderot, Sorbonne Paris Cit\'{e},\\ Institut Langevin, UMR 7587, 1 rue Jussieu, F-75005 Paris, France}
\author{Michel Campillo}
\affiliation{ISTerre, Universit\'{e} Joseph Fourier, Maison des G\'{e}osciences, BP 53, F-38041 Grenoble, France}
\author{Alexandre Aubry}\email{alexandre.aubry@espci.fr}
\affiliation{ESPCI Paris, PSL Research University, CNRS, Univ Paris Diderot, Sorbonne Paris Cit\'{e},\\ Institut Langevin, UMR 7587, 1 rue Jussieu, F-75005 Paris, France}

\begin{abstract}
Multiple scattering of seismic waves is often seen as a nightmare for conventional migration techniques that generally rely on a ballistic or a single scattering assumption. In heterogeneous areas such as volcanoes, the multiple scattering contribution limits the imaging-depth to one scattering mean free path, the mean distance between two successive scattering events for body waves. In this Letter, we propose a matrix approach of passive seismic imaging that pushes back this fundamental limit by making an efficient use of scattered body waves drowned into a noisy seismic coda. As a proof-of-concept, the case of the Erebus volcano in Antarctica is considered. The Green's functions between a set of geophones placed on top of the volcano are first retrieved by the cross-correlation of coda waves induced by multiple icequakes. This set of impulse responses forms a reflection matrix. By combining a matrix discrimination of singly-scattered waves with iterative time reversal, we are able to push back the multiple scattering limit beyond 10 scattering mean free paths. The matrix approach reveals the internal structure of the Erebus volcano: A chimney-shaped structure at shallow depths, a magma reservoir at 2500 m and several cavities at sea level and below it. The matrix approach paves the way towards a greatly improved monitoring of volcanic structures at depth. Beyond this specific case, the matrix approach of seismic imaging can generally be applied to all scales and areas where multiple scattering events undergone by body waves prevent in-depth imaging of the Earth's crust.
\end{abstract}

\maketitle

%
%

%


%
%
%
%
\section{Introduction}

{Conventional seismic modeling, imaging, and inversion methods generally apply to direct waves in seismic records~\citep{Yilmaz}. However, in heterogeneous areas, the {coherent} components of the wave-field may vanish, and the seismic record entails a long coda due to scattering undergone by seismic waves in the Earth's crust. Albeit difficult to interpret via conventional approaches, the information contained in the coda is particularly rich. The challenge is to extract relevant information from such a seemingly incoherent signal. Cross-correlation of coda waves or seismic noise received at two stations was shown to be an important breakthrough in geophysics~\citep{campillo,larose}. {Under appropriate wave-field conditions, coda cross-correlation converges towards} the Green's function between receiving stations~\citep{weaver,derode2,wapenaar,snieder}, as if one of them had been used as a source, thus paving the way to passive imaging.}

{In that process, multiple scattering of waves helps approaching energy equipartition~\citep{hennino,margerin} which is necessary to fully retrieve the exact Green's function. This seems paradoxical because multiple scattering is {detrimental to} imaging: it becomes more and more difficult to distinguish anything at depths larger than the scattering mean-free path $\ell_s$, the mean distance between two successive scattering events. Beyond that depth, the wave actually loses its coherence and the memory of its initial direction. In reality, multiple scattering does help {to retrieve} the main features of the Green's function passively, but subsequent imaging techniques ultimately apply to the ballistic (\textit{i.e.}, unscattered) or singly-scattered component of the estimated Green's functions. \cora{Although a multiply scattered wavefield is a prerequisite to passively recovering Green's functions through cross-correlations, ultimately passive imaging suffers from the same limitations as classical active imaging: it fails if multiple scattering is too strong.}

\cora{Yet new perspectives to overcome this problem were opened by the advent of multi-element arrays with controllable emitters and/or receivers}, both in ultrasound imaging~\citep{aubry,aubry2,shahjahan} and in optical microscopy~\citep{badon2016,badon_optica}. In a linear and time-invariant system with an array of $M$ independent emitters/receivers, the propagation of waves can be described by a matrix approach. All relevant information is contained in the $M \times M$ reflection matrix {composed of} the set of inter-element impulse responses. In this context, recent academic studies proposed a solution based on random matrix properties to overcome multiple scattering by separating the single and multiple scattering contributions in the reflection matrix. From a physical point of view, it was done by taking advantage of the memory effect in the far-field~\citep{shahjahan} or, equivalently, by a confocal discrimination between single and multiple scattering in the focal plane~\citep{badon2016}.}

\cora{Inspired by these previous works, this paper aims to develop a similar matrix approach for seismic wave imaging in strongly scattering environments. Arrays of geophones are commonly used to probe the subsoil, hence matrix formalism can be particularly appropriate, as is the case for ultrasound and optical imaging.  As a proof-of-concept for seismic imaging,} we here consider the case of the Erebus volcano in Antarctica. Volcanoes are among the most challenging media for seismic imaging {given their} highly localized and abrupt variations {in} physical parameters, extreme landforms, extensive fractures {and the additional presence} of magma and other fluids. In that respect, the case of the Erebus volcano is extreme since the scattering mean free path $\ell_s$ for body waves is smaller than their wavelength $\lambda$ in the 1-4 Hz bandwidth~\citep{Chaput2015}. Imaging the volcano at large depths with conventional techniques fails, since multiple scattering {increasingly} dominates as depth becomes significantly larger than $\ell_s$. Strikingly, we will show that our matrix approach allows {us} to push back the multiple scattering limit to a depth of 9000 m, or roughly $12 \ell_s$ at a frequency $f=2.6$ Hz, for which $\ell_s \sim 750$ m~\citep{Chaput2015}.

To do so, we take advantage of a high-density network of 76 seismographs that was deployed on {the upper plateau} of the volcano {as part of the TOMO-Erebus project (2007-2009)}. The coda generated by shallow icequakes~\citep{Knox} is used to retrieve the {vertical component of the} Green's functions between the geophones. The associated reflection matrix is then investigated for imaging purposes. \cora{As a whole, the process we present in this paper can be analyzed as a combination of 5 building blocks (see Fig.~S1 in supporting information):
\begin{itemize}
\item (B1) Any reflection matrix can be written as a sum of three contributions: ballistic, single-scattering, multiple scattering.
\item (B2) Based on a rough estimate of velocity $c$, the Green's functions between the $M$ actual geophones can be transformed in order to mimic a set of $N$ virtual geophones, located in a plane at an arbitrary depth $z$ below the surface. To that end, focusing is performed both at emission and reception by means of simple matrix operations. It yields a new reflection matrix, each element of which is the total response (\textit{i.e.}, ballistic, single and multiple scattering) between two virtual geophones~\citep{robert2,badon2016,wapenaar2}.
\item (B3) An input-output analysis, which will be referred to as confocal filtering, allows for the removal of most of the multiple scattering contribution in the new reflection matrix~\citep{badon2016}.
\item (B4) Iterative time reversal \citep{prada,prada2} is applied to overcome the residual multiple scattering contribution as well as the aberration effects induced by the scattering medium itself, and detect possible reflectors.
\item (B5) A statistical analysis of the matrix singular values permits attribution of a likelihood index to each detected structure \citep{aubry2,aubry_wrmc}.
\end{itemize}
As a result of these five steps, an image of the internal structure of the Erebus volcano is obtained. While conventional imaging methods lead to a speckled image due to multiple scattering, the matrix approach developed here manages to detect internal structures in a reliable way, and reveal a chimney-shaped structure feeding the lava lake that bifurcates sharply towards the north-west and then seems to set centrally in a shallow magma chamber near 2500 m elevation. Some particular features also emerge from the sea level to 5000 m below it. A structural interpretation of the obtained image is finally built on the existing literature about Erebus.}

\section{Response matrix of geophones}

\begin{figure}[htbp]
 \centering
 \includegraphics[width=14cm]{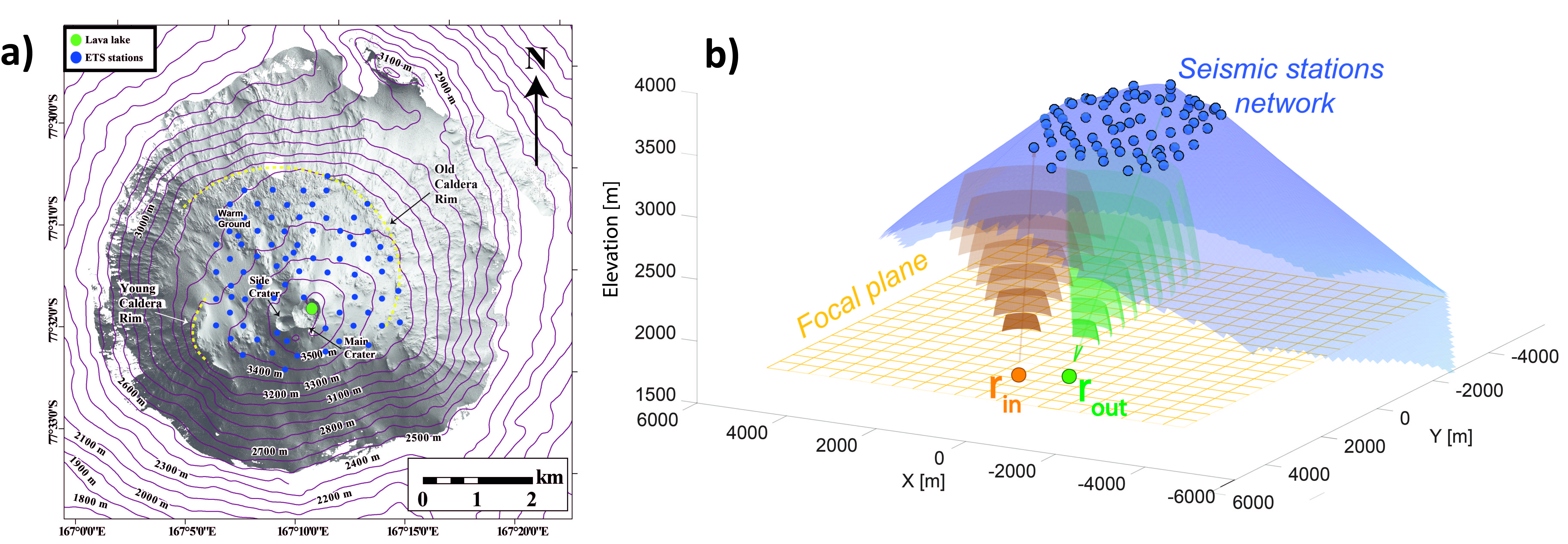}
 \caption{a) Map of the Erebus volcano upper edifice with some principal geographic and geological features and showing the locations of the ETS Erebus stations. b) Adaptive focusing at emission and reception on two points $\mathbf{r_{in}}$ and  $\mathbf{r_{out}}$ of the focal plane ($z=ct/2$) yields the Green's function between virtual geophones placed at these two points. The same operation is repeated for any couple of points in the focal plane and yields the focal reflection matrix $\mathbf{R}$ [Eq.~(\ref{GRG})].}
 \label{fig1}
  \end{figure}
In this study, we take advantage of inter-station correlations of ice-quake coda recorded on a large temporary short-period deployment (referred to as the ETS network). \cora{Previous studies have identified several distinct mechanisms for short period discrete seismicity, namely Strombolian gas slug decompression eruptions from a centralized convecting lava lake~\citep{Chaput2012,Knox2018}, and a plethora of randomly distributed impulsive icequake events on Erebus volcano's snow and ice cap~\citep{Knox}. Though similar in both temporal and spectral character, eruptive events were systematically separated from icequake events in early studies~\citep{Chaput2012} using a joint seismic and infrasound matched filter. Icequake events were subsequently detected and roughly located using the Antelope database management package, using an STA/LTA trigger coupled with grid search waveform stacking for a simple velocity model. Spatial resolution of detected events, as noted in Ref.~\cite{Knox}, is on the order of 10-100 m based on the number of recording stations and the location of the icequake (inside or outside the network). The coda of both types of events were shown to satisfy theoretical precursors of equipartition~\citep{Chaput2015}, notably, modal partition stabilization~\citep{hennino}, and coherent backscattering~\citep{Larose2006}. Further studies using these icequake events also noted that exact source locations were unnecessary for Green's function reconstruction, given that highly symmetric coda correlation functions could be obtained with a correct choice of coda windows, while using a severally azimuthally polarized subset of the icequakes~\citep{Chaput2015,chaput2016}. These points together, along with numerically validated Bayesian optimization of coda windows as proposed by Ref.~\cite{chaput2016}, form a set of corroborating arguments that point to an accurate convergence of icequake coda correlations to the true Green's function between station pairs.}

Figure \ref{fig1}a shows the location of the ETS instruments \citep{Zandomeneghi,Chaput2012}. The approximately 5 km diameter summit plateau, cone, and crater region of the Erebus volcano is a caldera-filling, highly heterogeneous, assemblage of bomb deposits, buried intrusions, geothermal features, gas-rich magma, permafrost, and lava flows. The volcano hosts an approximately 500 m diameter main crater that {hosts} a long-lived phonolitic lava lake sustained by a significant near-summit magmatic system \citep{Zandomeneghi2013}. The 76 seismographs, spaced 300 to 500m apart, were deployed over a roughly 4 by 4 km area surrounding the summit crater and lava lake. Each geophone is denoted by an index $i$ and its position $\mathbf{s_i}$. We make use here of 3318 randomly located icequake events identified by Ref.~\cite{Knox} for a period of 1 month in 2008-2009 \citep{Zandomeneghi}. We performed inter-station correlations of ice-quake coda using the vertical component of the recorded wave-fields. The method used for the convergence of the coda cross-correlation towards inter-station Green's functions has been detailed by {Ref.~\cite{chaput2016}, and relies on a Bayesian optimization scheme to select ideal combinations of coda windows at each icequake.} The impulse response between stations $i$ and $j$ is noted $h_{ij}(\tau)$, with $\tau$ the time lag. The set of impulse responses form a time-dependent response matrix $\mathbf{H}(\tau)$. The Green's functions are estimated over a time length of 15 s, with a 20 Hz sampling frequency. The central frequency $f_0$ is 2.6 Hz and the {$-3$ dB} frequency bandwidth is 1 Hz. 

The impulse responses exhibit several direct arrivals that have already been investigated by {Refs.~\cite{Chaput2015,chaput2016}}. Rayleigh waves are well resolved, with an apparent velocity ranging between 1000 and 1300 m/s. Ballistic waves, likely direct inter-station S-wave and P-wave, arrive before the Rayleigh wave at apparent velocities between 1500 m/s and 3000 m/s. This range of body wave velocities is consistent with a previous seismic tomography study~\citep{Zandomeneghi2013} from which a mean P-wave velocity of 2200 m/s has been measured in the near-surface region of the edifice. \cora{In the following, we will use a homogeneous wave velocity model with $c= 2200$ m/s as the reference bulk wave velocity.} This choice will be also validated \textit{a posteriori} by a minimization of the aberration effects in the reflection matrix (see Sec.~\ref{focal}). 

In the present study, we are not interested in the ballistic component of the wave-field {but rather in} its scattered contribution due to reflections by the internal structure of the volcano. {These vertical echoes are mainly associated with P-waves since only the vertical component of the Green's functions is considered in this study.} The scattered wave-field {consists of} two parts: (i) a single-scattering contribution which can be taken advantage of for 3D imaging, because there is a direct first-order relation between the arrival time $\tau$ of singly-scattered echoes and the distance $d$ between sensors and scatterers, $\tau = 2d/c$; (ii) a multiple-scattering contribution for which the time-space relation does not hold, which is a {hurdle} for imaging. On the one hand, a time-resolved analysis of $\mathbf{H}$ is needed in order to enhance the single-scattering component compared to the multiple scattering contribution. On the other hand, a Fourier analysis of the impulse responses is required for a matrix description of wave propagation. 

To fulfill these two constraints, a short-time Fourier analysis of the impulse response matrix $\mathbf{H}$ is achieved. The impulse responses $h_{ij}(\tau)$ are truncated into successive time windows of length $\Delta t$: \cora{$k_{ij}(t, \tau) =h_{ij}(t+\tau) W(\tau)$} with $W(\tau) = 1$ for $\tau \in [-\Delta t/2, \Delta t/2]$, $W(\tau) = 0$ elsewhere. \cor{$t$ is the central time of each temporal window.} The value of $\Delta t=1$ s is chosen so that signals associated with the same scattering event(s) within the medium arrive in the same time window\cor{~\citep{aubry_wrmc}}. For each value of time $t$, the coefficients $k_{ij}(t,\tau)$ form a matrix $\mathbf{K}(t,\tau)$.  A Fourier analysis is then achieved by means of a discrete Fourier transform (DFT) and gives a set of impulse responses $k_{ij}(t,f)$ at time $t$ and frequency $f$. In the following, we will mainly consider the response matrix $\mathbf{K}(t,f_0)$ at the central frequency $f_0=2.6$ Hz. 

\cora{As previously discussed, any reflection matrix can be decomposed as the sum of a ballistic (B) contribution, a single (S) scattering contribution and a multiple (M) scattering contribution. In the case of $\mathbf{K}(t,f)$, this is written:
\begin{equation}
\mathbf{K}(t,f)=\mathbf{K_B}(t,f)+\mathbf{K_S}(t,f)+\mathbf{K_M}(t,f),
\end{equation}
or, in terms of matrix coefficients,
\begin{equation}
k_{ij}(t,f)=k^{(B)}_{ij}(t,f)+k^{(S)}_{ij}(t,f)+k^{(M)}_{ij}(t,f).
\end{equation}
At any specific time $t$, $\mathbf{K_S}(t,f)$ contains singly-scattered echoes associated with reflectors contained in the \textit{isochronous volume} \textit{i.e.}, the ensemble of points that contribute to the back-scattered signal at a given time. 
It is formed by the superposition of all ellipses whose foci are elements $i$ and $j$, for $(i,j) \in [1;M]$. In a far-field configuration, the isochronous volume can be approximated to a slab of thickness $\Delta r=c \Delta t$, centered at a distance $z = ct/2$ from the array of geophones and parallel to it.} The multiple scattering contribution $\mathbf{K_M}(t,f)$ corresponds to the sum of partial waves associated with multiple scattering paths that occur at shallower depths and whose length belongs to the interval $[z-c \Delta t/2;z+c \Delta t/2]$.
 
Previous works in ultrasound imaging have investigated the input-output correlation properties of the reflection matrix in order to discriminate single and multiple scattering~\citep{aubry,aubry2}. While the latter contribution displays a random feature, the former contribution displays a deterministic coherence related to the memory effect \citep{shahjahan}. {The original multiple scattering filter in ultrasound imaging has been developed for a regular array of sensors under a paraxial approximation and neglecting wave-front distortions related to an inhomogeneous wave velocity background~\citep{aubry,aubry2}. On the contrary, the geophones placed on top of Erebus have irregular positions relative to one another.} Moreover, the paraxial approximation is everything but true at shallow depths and wave velocity tomography maps show strong variations at the top of Erebus volcano~\citep{Zandomeneghi2013}. \cora{Hence, though we are dealing with seismic waves whose mechanical nature is similar to ultrasound, the discrimination between single and multiple scattering must be performed in a different manner than what has been done in the context of ultrasonic detection. Instead, we adapt a recent extension of the matrix approach to optical imaging ~\citep{badon2016}. In the next section we explain how the reflection matrix is transformed by emission and reception focusing, before discriminating single and multiple scattering.}

\section{\label{focal}Focal reflection matrix}

\cora{In order to emulate a set of virtual sources/receivers located below the surface at an arbitrary depth $z$, coherent beamforming can be achieved at emission and reception.} If the geophones were active sensors, it would consist in focusing the transmitted wave at the desired point by applying the appropriate time delay to each geophone. In the reception mode, the same delays would be applied to the received signals before being summed. Single scattering signals coming from a scatterer located at the focus would add up coherently, whereas the summation would be expected to be incoherent for multiple scattering signals arriving at the same time. 

The matrix formalism allows to do all these operations in post-processing. To that aim, we first have to map, at each time, the focal plane with a 2D set of $N^2$ focusing points along the $x$ and $y$ directions (see Figure~\ref{fig1}). The spatial sampling period should be chosen smaller than the typical size $\delta$ of a resolution cell. {$\delta$ is determined by the effective numerical aperture, $\sin \theta$, of the geophone array at each depth $z$, such that
\begin{equation}
\delta \sim \frac{\lambda}{2 \sin \theta} \mbox{, with }\tan \theta=\frac{D}{2z} 
\end{equation}
\cora{where $\lambda=c/f_0$ is the wavelength, assuming $c$ is known. $D$ is the typical size of the geophones' network.} At shallow depth, $\delta$ is limited by diffraction: $\delta \sim \lambda/2$. The number $N^2$ of focusing points in the focal plane is chosen such that the \textit{field-of-view} $N \delta$ (i.e., the transverse size of the focal plane) coincides with \cor{twice the geophones' array size at shallow depth, hence we take $N\lambda/2 \sim 2 D $. Here $D=4$ km,} so $N$ is fixed to be 40. In the far-field ($z > > D$), the resolution cell $\delta$ scales as $\lambda z/D$ and the field-of-view increases linearly with $z$.}

{The second step consists in defining a Green's matrix $\mathbf{G}$ that describes the ballistic propagation of seismic waves from the geophones' network to the focal plane. As the scattered wave-field mainly contains P-waves, an homogeneous {acoustic} model is considered with a wave velocity equal to $c$.} $\mathbf{G}$ gathers all the free-space causal Green's functions $G(\mathbf{s_i},\mathbf{r_l})$ between each geophone at point $\mathbf{s_i}$ and each focal point $\mathbf{r_l}$:
\begin{equation}
G(\mathbf{s_i},\mathbf{r_l})=\frac{\textrm{e}^{-\textrm{i}\frac{2\pi}{\lambda} \left\|\mathbf{r_l}-\mathbf{s_i} \right\|}}{4 \pi \left\|\mathbf{r_l}-\mathbf{s_i}\right\|} 
\end{equation}
The final step consists in applying the double focusing operation to $\mathbf{K}$ to obtain the reflection matrix $\mathbf{R}$ in the focal plane. \cora{Based on the Kirchoff-Helmholtz integral, such a focusing operation is standard in exploration seismology and referred to as \textit{redatuming}\citep{Berkhout,Berryhill,Berkhout2}. Under a matrix formalism, it can be written as follows,
\begin{equation}
\mathbf{R}=\mathbf{\partial_n G^{\dag}} \times \mathbf{K} \times \mathbf{\partial_n G^*} 
\label{GRG}
\end{equation}}
where the symbols $\dag$ and $*$ stands for transpose conjugate and conjugate. \cora{The matrices $\partial_n\mathbf{G^{\dag}}$ and $\partial_n\mathbf{G^{*}}$ contain the normal derivatives of the anti-causal Green's functions between the geophones and the focal points, i.e the wave-fronts that should be applied at emission and reception in order to project the reflection matrix in the focal plane (Figure~\ref{fig1}).} The coefficients $R(\mathbf{r_{in}},\mathbf{r_{out}} )$ of $\mathbf{R}$ is the impulse response between a virtual source at point $\mathbf{r_{in}}=(x_{in},y_{in},z)$ and a virtual detector at $\mathbf{r_{out}}=(x_{out},y_{out},z)$ at depth $z=ct/2$ \citep{Aubry2007,robert2}. The effective size of \cora{each} virtual geophone is $\delta$ in the transverse direction and $c\Delta t /2$ in the axial direction.  
\begin{figure}[htbp]
 \centering
 \includegraphics[width=12cm]{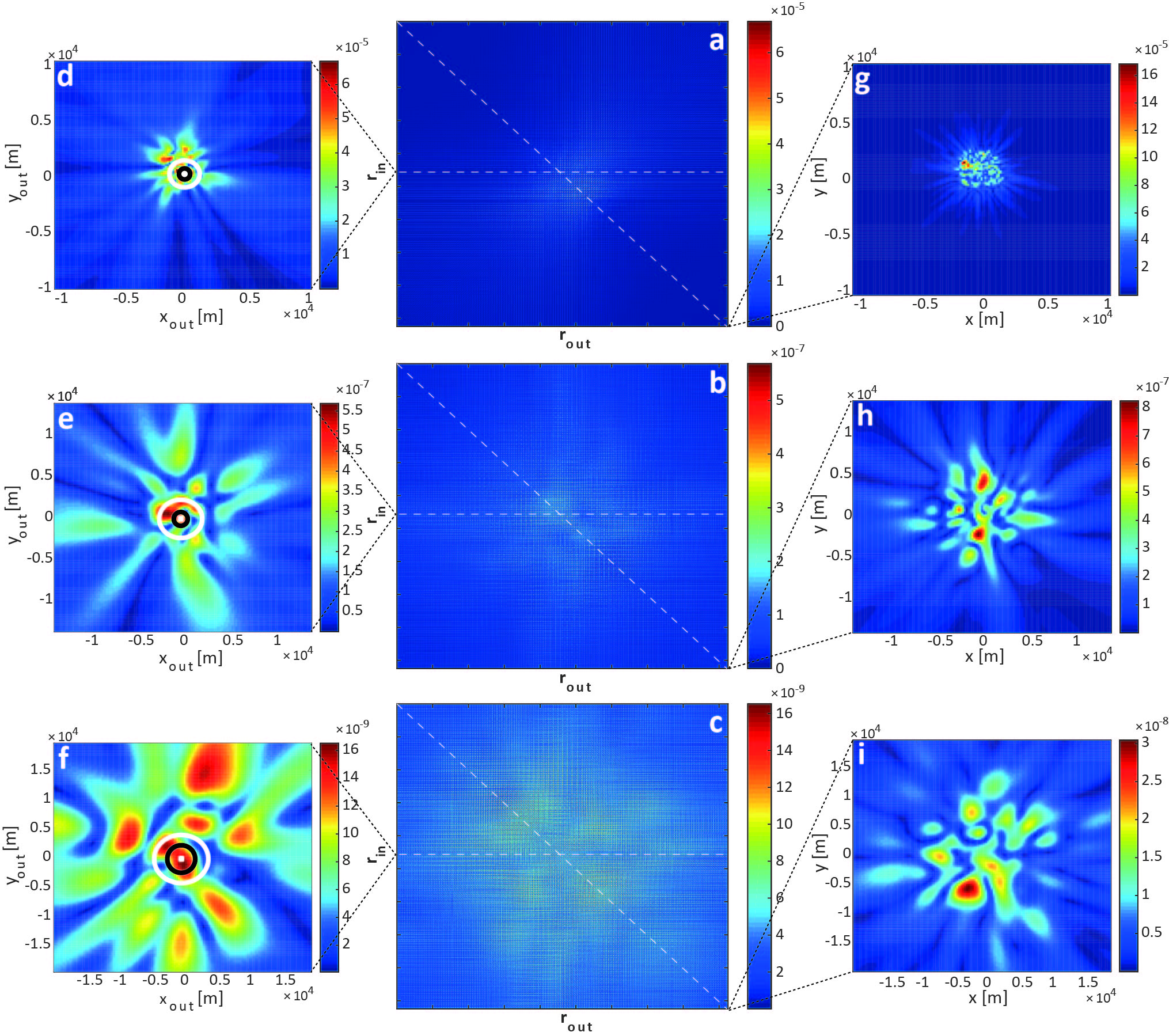}
 \caption{(a,b,c) Reflection matrices $\mathbf{R}$ in the focal plane basis recorded at time $t=0.25$, 5 and 12.5 s. (d,e,f) The central line of these reflection matrices is reshaped in 2D and yields the focal spot for a central input focusing point $\mathbf{r_{in}}=\mathbf{0}$. The black circle represents the ideal focal spot of radius $\delta$ that would be obtained in absence of aberration. The white circle line accounts for the typical size $\ell_c$ of the aberrated focal spot. (g,h,i) The diagonal of these reflection matrices is reshaped in 2D and yields the confocal image of each focal plane. Note that the field-of-view scales with the size of the resolution cell predicted by diffraction theory such that each focal plane contains a constant number of points. {In each panel, the modulus of matrix coefficients is displayed in arbitrary units with a linear color scale}.}
 \label{fig3}
  \end{figure}

\begin{figure}[htbp]
 \centering
 \includegraphics[width=12cm]{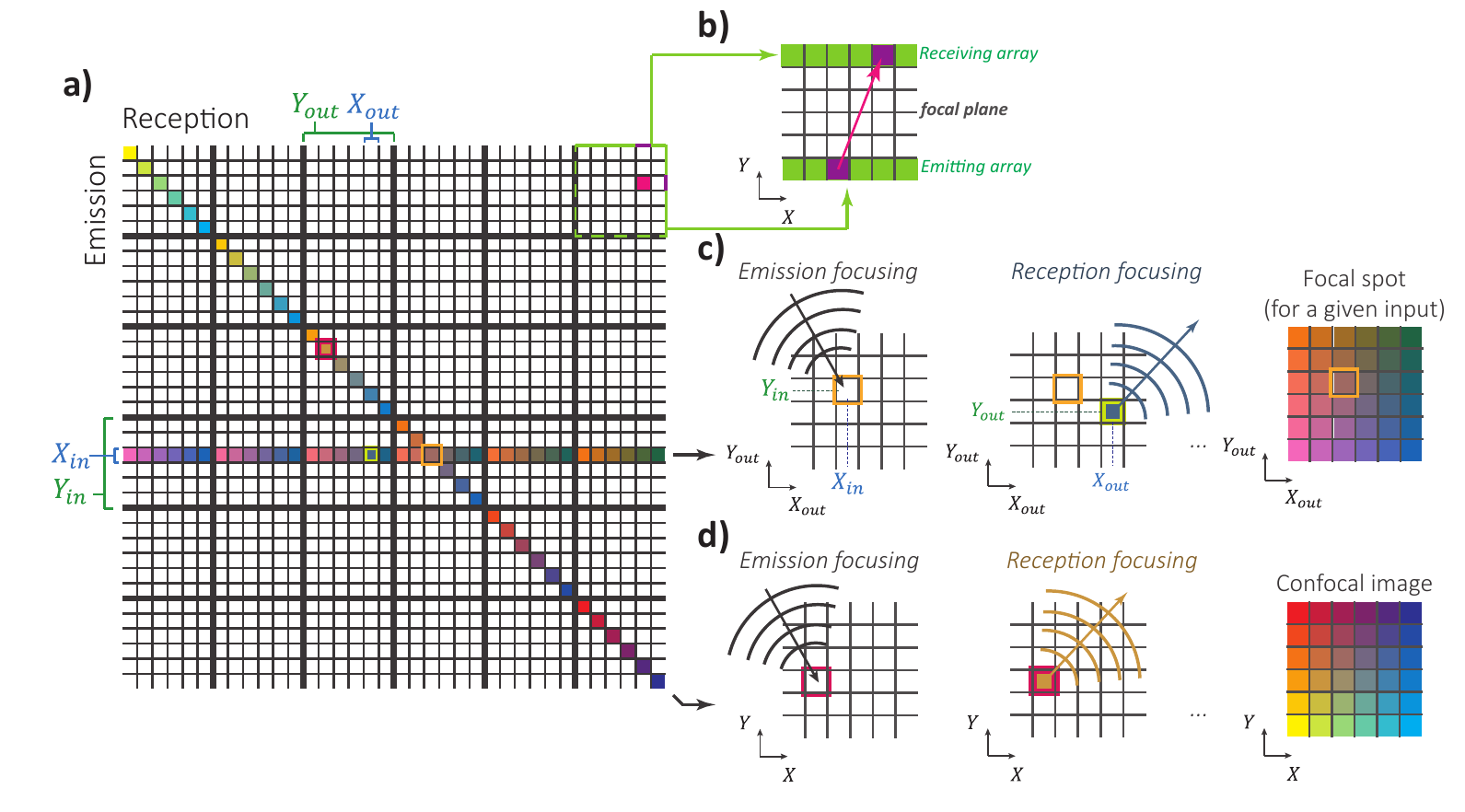}
 \caption{Structure of the focal reflection matrix $\mathbf{R}$. a) Every line or column corresponds to the reflected wave-field for an input or output focusing point, $\mathbf{r_{in}}$ or $\mathbf{r_{out}}$, respectively. b) The 2D representation of $\mathbf{R}$ corresponds to a set of block matrices that represents the reflection matrix between two lines of virtual geophones in the focal plane. c) A line in the reflection matrix provides the backscattered wave-field backpropagated in the focal plane for an input focusing point $\mathbf{r_{in}}$. d) The diagonal of $\mathbf{R}$ provides the reflected wave-field for identical input and output focusing points. It yields a confocal image of the volcano at depth $z={c t}/{2}$.}
 \label{fig2}
  \end{figure}

Figures \ref{fig3}a, b and c display the focal reflection matrix at three different times of flight: $t=$ 0.25 s, 5 s and 12.5 s, respectively. Since the focal plane is bi-dimensional, $\mathbf{R}$ has a four-dimension structure: $R(\mathbf{r_{in}},\mathbf{r_{out}})=R(x_{in}, y_{in}, x_{out},y_{out})$. $\mathbf{R}$ is thus concatenated in 2D as a set of block matrices to be represented graphically [see Figure~\ref{fig2}a]. In such a representation, every sub-matrix of $\mathbf{R}$ corresponds to a specific couple $(y_{in},y_{out} )$, whereas every element in the given sub-matrix corresponds to a specific couple $(x_{in},x_{out} )$ [Figure~\ref{fig2}b]. Each coefficient $R(\mathbf{r_{in}}, \mathbf{r_{out}})$ corresponds to the complex amplitude of the echoes coming from the point $\mathbf{r_{out}}$ in the focal plane when focusing at point $\mathbf{r_{in}}$ (or conversely since $\mathbf{R}$ is a symmetric matrix due to spatial reciprocity). Therefore, a line of the reflection matrix yields the response of the medium all across the focal plane for a specific input focusing point. It may be reshaped in two dimensions to visualize the corresponding reflected wave-fields in the focal plane (Figure~\ref{fig2}b). Figures \ref{fig3}d, e and f show the 2D reshaped central line of the matrices $\mathbf{R}$ displayed in Figures \ref{fig3}a, b and c, respectively and in these examples, the input focusing point is at the center of the field-of-view. 

At small times of flight ($t=0.25$ s) or equivalently at shallow depths {($z \sim 275$ m}), single scattering is predominant since the corresponding focal depth is shallower than $\ell_s$. Not surprisingly, $\mathbf{R}$ is close to {being} a diagonal matrix (Figure~\ref{fig3}a). Most of the incident energy reaches the input focusing point and most of the reflected echoes are collected nearby, which is a characteristic feature of singly-scattered echoes. Nevertheless, as illustrated by Figure~\ref{fig3}d, the characteristic size of the focal spot is larger than the resolution cell $\delta$ predicted by diffraction theory. The size of the focal spot is given by the coherence length \cora{$\ell_c$} of the reflected wave-field back-propagated in the focal plane. At time $t=0.25$ s, $\ell_c$ is twice the theoretical resolution cell $\delta$. The inhomogeneity of the bulk wave velocity implies wave-front distortions and an energy spreading over off-diagonal elements of $\mathbf{R}$ in Figure~\ref{fig3}a. {The blurring of the reflection matrix around the diagonal is a manifestation of the aberrations undergone by the incident and reflected wave-fronts during their travel from the geophone array to the focal plane. This property has been taken advantage of to validate \textit{a posteriori} our initial choice for the mean wave velocity. The selected value, $c=2200$ m/s, is found to minimize the ratio $\ell_c/\lambda$. Such an optimization is important since it has an impact on both the transverse resolution and the stretching in depth of the synthesized image.}

At larger depths ($\ell_s < z < 10\ell_s$), we are in an intermediate regime where single and multiple scattering coexist. Figures \ref{fig3}b and e display an example of reflection matrix and its central focal spot in this intermediary regime ($t=5$ s, $z\sim 5.5$ km). An aberrated focal spot can hardly be observed in the vicinity of the input focusing point (Figure~\ref{fig3}e). Aberration can be assessed by the ratio $\ell_c/\delta \sim 3$, which is stronger than at shallow depths (Figure~\ref{fig3}d). Otherwise, beyond the central area ($ \| \mathbf{r_{in}} - \mathbf{r_{out}} \|>\ell_c$), the reflected wave-field also shows much higher side lobes that can be accounted for by multiple scattering events occurring at shallower depths. With regards to the reflection matrix (Figure~\ref{fig3}b), the multiple scattering contribution manifests itself as a spreading of the back-scattered energy over its off-diagonal elements.  

At last, beyond a depth of $10\ell_s$, we are in a fully developed multiple scattering regime. Figures~\ref{fig3}c and f show a reflexion matrix and its 2D reshaped central line in this regime ($t=12.5$ s, $z\sim 13.75$ km). Unlike previous cases, the backscattered wave-field displays a random feature with hardly any relationship to the original input location (Figure~\ref{fig3}f). At large travel times, backscattered waves have actually undergone numerous scattering events and followed complex trajectories. Not surprisingly, multiple scattering yield a seemingly random reflection matrix (Figure~\ref{fig3}c). 

The results displayed in Figure\ref{fig3} have shown how the reflection matrix gives a quantitative information about the aberration level and the single-to-multiple scattering ratio. In the next sections, we will show how the internal structure of the volcano can be imaged from the reflection matrix. 

\section{In-depth imaging of Erebus volcano}

\subsection{Confocal imaging}

A direct way to build {a map of reflectivity for each focal plane, hence a 3D image,} is to consider the diagonal of the reflection matrix $\mathbf{R}$. Each diagonal element $R(\mathbf{r_{in}},\mathbf{r_{in}})$ corresponds to the recorded echo amplitude when focused beamforming is performed both at emission and reception on the same point $\mathbf{r_{in}}$. Repeating this operation for each point within the field-of-view gives an image, each pixel of which codes the reflectivity of a specific point $\mathbf{r_{in}}$. Confocal imaging is actually the underlying principle of ultrasound imaging. It is based on a single-scattering assumption and is extremely sensitive to aberration issues.

Figures \ref{fig3}g, h and i display the 2D confocal images built from the diagonal of the reflection matrices displayed in Figures 3a, b and c at time $t=0.5$ s, 5 s and 10 s. At shallow depth (Figure~\ref{fig3}g), one could have expected the confocal image to provide a satisfying image of the Erebus volcano. However, the short propagation distance and the aberration level (Figure~\ref{fig3}d) preclude a clear distinction of inner structures of the volcano {beyond roughly 1 km depth. Note that Ref.~\cite{Chaput2012} already obtained a scattering image of Erebus in this depth range by means of a confocal imaging process. However, this work benefited from additional arrival incidence information provided by the full Green's tensor, itself estimated by correlations of Strombolian eruption coda.} At larger depths (Figures~\ref{fig3}h-i), multiple scattering predominates over single scattering. A random speckle image is obtained with, probably, no connection with the reflectivity of the volcano at the corresponding focal depth. This is confirmed by Figure~\ref{fig6}a that displays the 3D confocal image of Erebus. 

Confocal imaging here fails not only because of multiple scattering but also owing to aberrations. A strict confocal discrimination of single scattering events is actually too restrictive. Aberrations induced by the wave velocity inhomogeneities degrade the singly-scattered focal spots that spread over the off-diagonal elements of the reflection matrix. In the next section, we show how to discriminate more precisely the single and multiple scattering contributions for imaging purposes.
	
\subsection{\cora{Confocal filter}}

\begin{figure}[htbp]
 \centering
 \includegraphics[width=10cm]{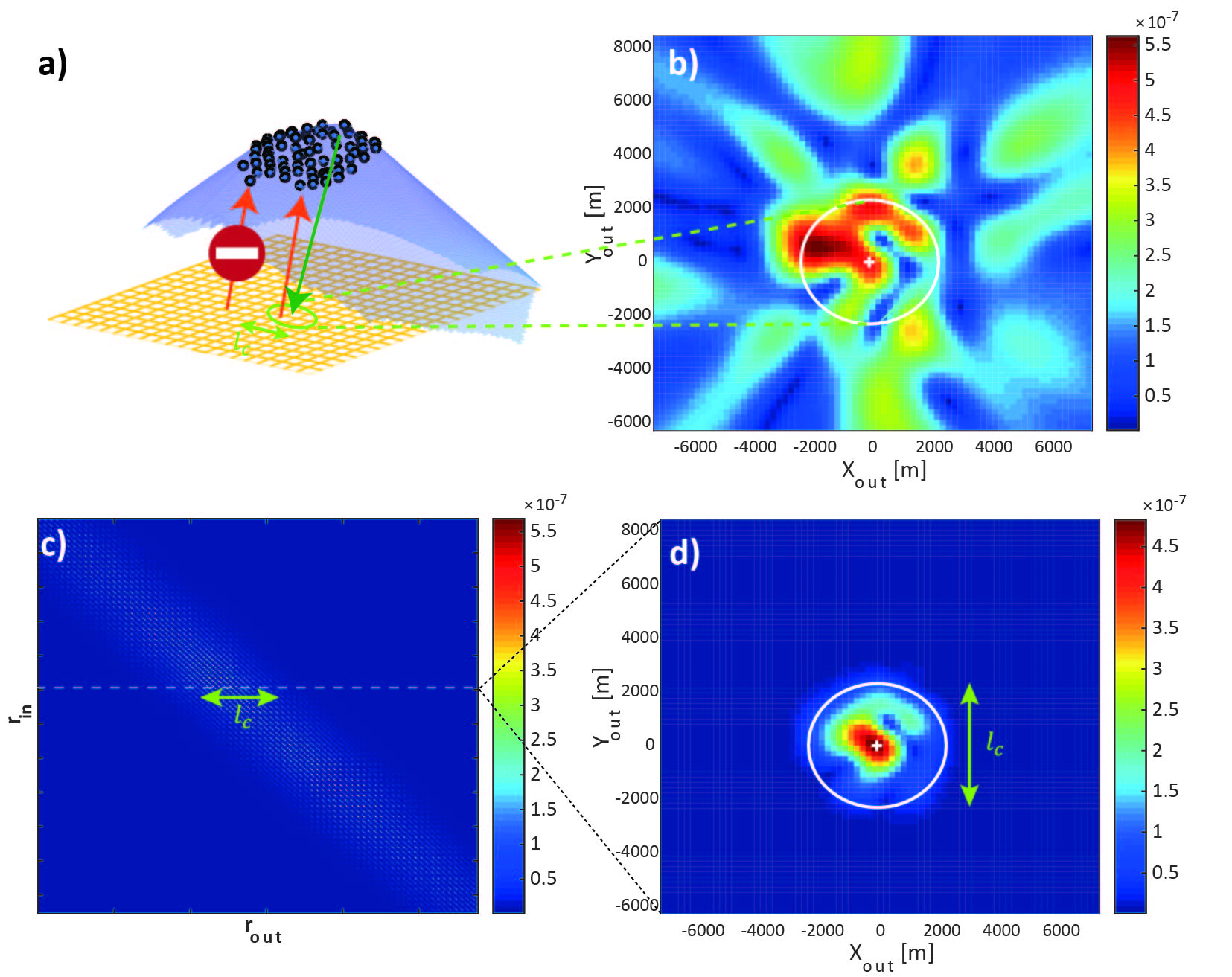}
 \caption{{Adaptive confocal filter of single scattering. a) Scheme of the confocal filtering operation described in Eq.~(\ref{filter}). (b) Reflected wave-field in the focal plane for an incident focusing wave at the origin of the focal plane at time $t=$5 s. c) Filtered reflection matrix $\mathbf{R_F}$ deduced from $\mathbf{R}$ displayed in Figure~\ref{fig3}b. [Eq.~(\ref{filter})].} {d) Adaptive confocal filter applied to the wave-field displayed in b.}}
 \label{fig4}
  \end{figure}

Recently, a new route has been proposed to extract singly-scattered echoes from a predominant multiple scattering background in presence of aberrations~\citep{badon2016}. {It consists in applying consecutively to the reflection matrix $\mathbf{R}$: (i) an adaptive confocal filter that we describe below; (ii) an iterative time reversal process that we will describe in Sec.~\ref{iterative}.} The adaptive confocal filter consists in weighting the coefficients $R(\mathbf{r_{in}},\mathbf{r_{out}})$ of the focal reflection matrix as a function of the distance $||\mathbf{r_{out}}-\mathbf{r_{in}}||$ between the virtual geophones. \cora{The characteristic size of the adaptive confocal filter matches the coherence length $\ell_c$ of the backscattered wave-field (Figure~\ref{fig4}a), that was introduced in Sec.~\ref{focal}. The filter is chosen to be Gaussian-shaped}
\begin{equation}
R_F(\mathbf{r_{in}},\mathbf{r_{out}})=R(\mathbf{r_{in}},\mathbf{r_{out}}) \times \exp \left [ - \frac{ \left \|  \mathbf{r_{in}}-\mathbf{r_{out}} \right \|^2}{2\ell_c^2 } \right ]
\label{filter}
\end{equation}
The effect of confocal filtering is illustrated in Figure~\ref{fig4}; it displays the filtered reflection matrix $\mathbf{R_F}$ (Figure~\ref{fig4}c) at time $t=5$ s. The original reflection matrix $\mathbf{R}$ is shown for comparison in Figure~\ref{fig3}b. Figures \ref{fig4}b-d display one line of  $\mathbf{R}$ and $\mathbf{R_F}$ reshaped in a 2D wave-field. They correspond to the reflected wave-field from the focal plane for an input focusing point at the origin. The confocal filter of Eq.~(\ref{filter}) tends to favor single scattering paths that are associated with input and output focusing points distant from less than one coherence length $\ell_c$. The further the input and output points are apart, the likelier the resulting signal is to contain multiple scattering, and the more it is accordingly penalized through the filter. As a consequence, even though $\mathbf{R}$ is clearly dominated by multiple scattering, $\mathbf{R_F}$ indeed displays a single scattering feature with most of the back-scattered energy gathered by close-diagonal elements (Figure~\ref{fig4}c) and much narrower focal spots {than the original matrix $\mathbf{R}$} (see \textit{e.g} the comparison between the central focal spots displayed Figures~\ref{fig4}b and d).

However, $\mathbf{R_F}$ still contains a residual multiple scattering component. Indeed, multiple scattering also contributes to the diagonal of $\mathbf{R}$ \cor{which is unaffected by the adaptive confocal filter in Eq.~\ref{filter}}. As a result, standard confocal imaging applied to $\mathbf{R_F}$ would also fail, resulting in the random speckle images displayed in Figures~\ref{fig3}h and i. Another operation is needed to filter the residual multiple scattering contribution. \cora{This is done using iterative time reversal~\citep{prada,prada2,badon2016}, presented in the next section.}
	
\subsection{\label{iterative}Iterative time reversal}

Iterative time reversal has originally been developed in acoustics. It initially consisted in insonifying a multi-target medium with a pulse wave, recording the back-scattered wave-field, and time-reversing it. Then, this time-reversed response was sent back into the medium and the whole process was iterated \citep{prada}. The procedure has been shown to converge towards a wave-front that selectively focuses on the strongest scatterer, even in presence of strong aberrations. Mathematically, this amounts to looking for the first invariant of the time reversal operator $\mathbf{K}\mathbf{K}^\dag$.

Iterative time reversal paved the way towards the DORT method (French acronym for Decomposition of the Time Reversal Operator), which takes advantage of the response matrix $\mathbf{K}$ associated with an array of transducers \cora{probing a homogeneous medium in which discrete scatterers are embedded~{\citep{prada,prada2}}.} It consists in computing the whole set of invariants of the time reversal operator $\mathbf{K}\mathbf{K}^\dag$ from its eigenvalue decomposition, or equivalently, from the singular value decomposition (SVD) of $\mathbf{K}$. Under a single scattering assumption, there is a one-to-one correspondence between each eigenstate of $\mathbf{K}$ and each scatterer embedded in the medium. Each singular value is proportional to the target reflectivity and the corresponding singular vector yields the wave-front that selectively focuses on each scatterer, even in presence of strong aberrations.

{In the present case}, the single scattering assumption is far from being verified and the one-to-one matching between scatterers and singular states of $\mathbf{K}$ is no longer true. Instead of working on the reflection matrix at the volcano surface, we consider it in the focal plane, and its \cora{\textit{confocally}} filtered version $\mathbf{R_F}$ is brought into play. Due to the confocal filter, the multiple scattering component has been severely reduced in $\mathbf{R_F}$. {Next, we rely on DORT for extracting singly-scattered echoes among the residual multiple scattering noise.} The SVD of $\mathbf{R_F}$ reads: 
\begin{equation}
\mathbf{R_F}=\mathbf{U\Lambda V}^\dag=\sum_{p=1}^{N^2} \lambda_p \mathbf{U_p} \mathbf{V_p}^\dag
\end{equation}
where $\mathbf{U}$ and $\mathbf{V}$ are unitary matrices whose columns $\mathbf{U_p}$ and $\mathbf{V_p}$ are the singular vectors. $\mathbf{\Lambda}$ is a real diagonal matrix whose elements $\lambda_1\geq\lambda_2\geq \cdots \geq\lambda_{N^2}\geq 0$ are the singular values. In the present case, the symmetry of $\mathbf{R_F}$ implies that $\mathbf{U}=\mathbf{V^*}$. \cora{As was mentioned before, each eigenstate of  $\mathbf{R_F}$ should be associated with one scatterer only, under a single scattering assumption. Note that, in the case of $\mathbf{R_F}$, this assumption refers to scattering trajectories starting and ending at the virtual transducers, not the actual geophones. }

Figure~\ref{fig5}a displays the singular values of $\mathbf{R_F}$ at time $t=5$ s. The first two singular values seem to emerge from a continuum of lower singular values associated with the residual multiple scattering noise in $\mathbf{R_F}$. The square norm $ \| U_p(\mathbf{r_{in}})\|^2$ of the corresponding singular vector coefficients are displayed in Figures~\ref{fig5}b and c. They show an energy focusing at specific locations. However, a systematic and quantitative analysis of the singular value spectrum is needed in order to assess whether the first two singular values are indeed associated with coherent reflectors at the focal depth or if they are false alarms due to speckle fluctuations. To that aim, a probabilistic approach based on random matrix theory is developed in the next section in order to discriminate potential artifacts~\citep{aubry2,aubry_wrmc,Shahjahan2017}.     

\subsection{\label{rely}{Likelihood index} of potential targets}
\begin{figure}[htbp]
 \centering
 \includegraphics[width=14cm]{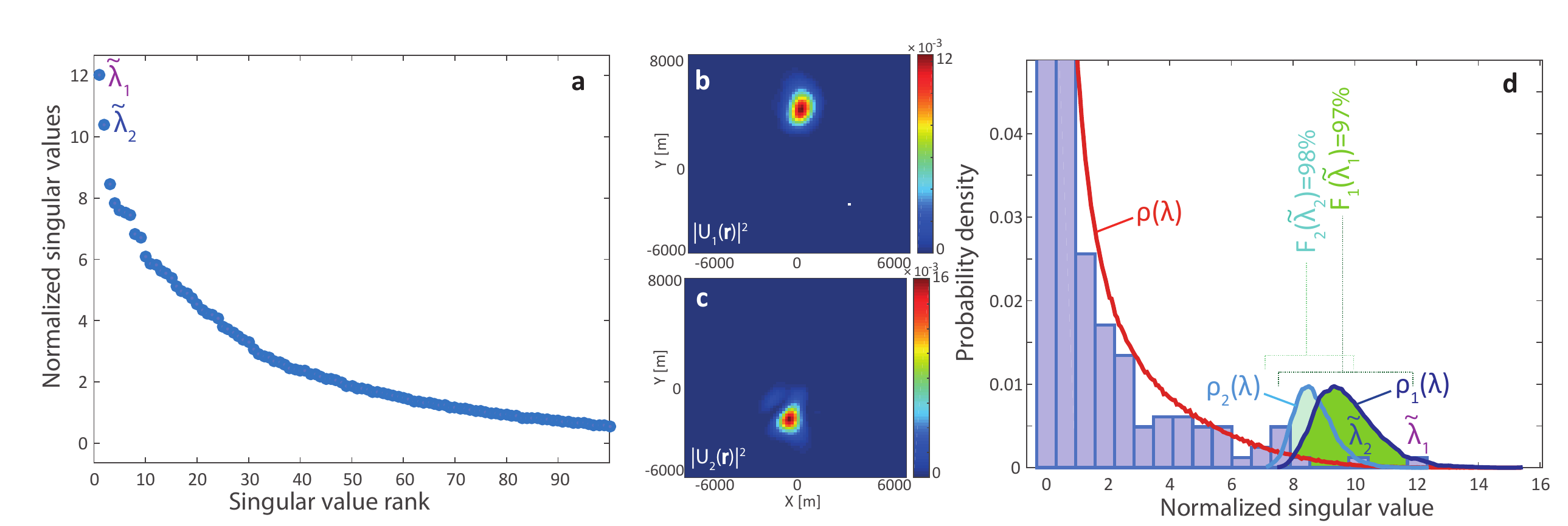}
 \caption{{Singular value decomposition of the reflection matrix $\mathbf{R_F}$ at time $t=5$ s (Figure~\ref{fig4}c). a) Normalized singular values $\tilde{\lambda}_p $ [Eq.~(\ref{norm})] displayed in decreasing order. b) Image associated with the first eigenspace of $\mathbf{R_F}$, $|U_1(\mathbf{r})|^2$. c) Image associated with the second eigenspace of $\mathbf{R_F}$, $|U_2(\mathbf{r})|^2$. d) Histogram of the experimental singular values compared with the probability density functions $\rho(\lambda)$ (red line), $\rho_1(\lambda)$ (dark blue line) and $\rho_2(\lambda)$ (light blue line) that would be obtained in a fully developed multiple scattering regime. The integration of $\rho_1(\lambda)$ (green shaded area) and $\rho_2(\lambda)$ (blue shaded area) from 0 to $\tilde{\lambda}_1 $ and $\tilde{\lambda}_2 $ [Eq.~(\ref{reliable})], respectively, yields the {likelihood index} of the corresponding eigenspaces.}
 }
 \label{fig5}
  \end{figure}

{Since the scattering medium is considered as one realization of a random process, assessing the reliability of the images obtained requires a statistical model for the probability density function of the singular values $\lambda_p$ in the multiple scattering regime. At any time $t$, by convention and for the sake of simplicity, the singular values $\lambda_p$ are normalized by their quadratic mean:
\begin{equation}
\label{norm}
\tilde{\lambda}_p=\frac{N\lambda_p}{\sqrt{\sum_{q=1}^{N^2}\lambda_q^2}}
\end{equation}
If the reflection matrix coefficients were independently and identically distributed random variables, the singular values would obey the quarter-circle law~\citep{marcenko}. However, because of the adaptive confocal filter applied to the raw reflection matrix $\mathbf{R}$, the variance of the matrix elements  $R_F(\mathbf{r_{in}},\mathbf{r_{out}})$ strongly depends on the distance $||\mathbf{r_{in}}-\mathbf{r_{out}}||$. Moreover, aberrations also imply short-range correlations over the coherence length $\ell_c$, which also impacts the distribution of singular values~\citep{sengupta}. For all these reasons, the histogram of normalized singular values displayed in Figure~\ref{fig5}d does not follow the quarter circle law. Therefore, in order to obtain a model for the probability density function $\rho(\lambda)$ in a target-less area, we numerically generated a set of random reflection matrices whose elements display the same statistical behavior (\textit{i.e.}, mean value, standard deviation and inter-element correlation) as the experimental matrix coefficients~\citep{aubry2011}. By repeating this operation over 500 realizations, the probability density function $\rho(\lambda)$ is estimated. Close agreement is found with the experimental histogram of the singular values, as shown in Figure~\ref{fig5}d.}

Once $\rho(\lambda)$ is known, a {likelihood index} can be associated with the first singular state. To that aim, the relevant quantity is the distribution function $F_1$ of the first singular value~\citep{aubry2,aubry_wrmc}: 
\begin{equation}
\label{reliable}
F_1(\lambda) =\mathbb{P} (\tilde{\lambda}_1 < \lambda)=\int_{0}^{\lambda} \rho_1(\lambda) d\lambda 
\end{equation}
with $\rho_1(\lambda)$ the probability density function of $\tilde{\lambda}_1$ [see Figure~\ref{fig5}d]. {The quantity $1-F_1(\tilde{\lambda}_1)$ is the probability that a pure multiple scattering speckle gives rise to a first singular value larger than the measured $\tilde{\lambda}_1$. Therefore, $F_1(\tilde{\lambda}_1)$ can be directly used as a likelihood index that the first singular state of $\mathbf{R_F}$ is actually associated with a coherent reflector (a \textit{target}) rather than a \textit{normal} multiple scattering speckle fluctuation.} For instance, at the specific time of flight $t=5$ s considered in Figure~\ref{fig5}, the first singular state is associated with a target with a high {likelihood index} of 97\%.  

{A likelihood index} can also be assigned to any eigenstate of rank $p>1$. To that aim, we need to determine the singular value distribution $\rho_p(\lambda)$ for a reflection matrix whose signal subspace is of rank $(p-1)$. Ref.~\cite{sengupta} has shown that the $(N^2-p)$ smallest singular values (linked to the noise subspace) exhibit the same distribution as singular values of a random matrix whose size is $(N^2 - p)\times N^2$. The corresponding distribution function $F_p(\lambda) $ can then be derived numerically following the same method as above. By way of example, the {likelihood index} of the second singular state in Figure~\ref{fig5} is of 98\%.  

\section{Three-dimensional imaging of the Erebus volcano}

\cora{Before building a 3D image of Erebus, let us recap the main steps of the procedure (see Fig.~S1 in supporting information):
\begin{itemize}
\item The Green's functions between the actual geophones have been recovered from cross-correlation of icequakes;
\item	The resulting matrix $\mathbf{H}$ has been submitted to a time-frequency analysis;
\item	Assuming a constant velocity, focused beamforming has been achieved in order to emulate a virtual array of sources and receivers, at any arbitrary depth;
\item A confocal filter, whose cutoff length is adapted to match the coherence length of the scattered field at each depth, is applied;
\item	The resulting matrix $\mathbf{R_F}$ undergoes an SVD in order to extract its strongest eigenstates, and the probability for these eigenstates to be due to an actual reflector is evaluated, based on random matrix theory.
\end{itemize}
So at this stage, we have all the necessary ingredients to build images and compare the result of the matrix approach to standard techniques.
}

{The iterative time reversal process described in Sec.~\ref{iterative} is applied at each time of flight. The images obtained from the first and second singular vectors of $\mathbf{R_F}$ at each ballistic depth are stacked into 3D images. The corresponding isosurface plots are displayed in Figure~\ref{fig6}b and c. The statistical analysis described in Sec.~\ref{rely} allows to assign a {likelihood index} to each detected scattering structure. {Note that only scattering structures associated with a likelihood index larger than 70\% are displayed. Inferior values mean that the corresponding singular state has a high probability of being associated with pure multiple scattering speckle}. Figure \ref{fig6}b displays the image provided by the first singular vector $\mathbf{U_1}$. Internal scattering structures of the volcano are imaged with a {likelihood index} higher than 90\% until an elevation of $-3000$ m, \textit{i.e.} a depth $z\sim9$ $l_s$. Figure \ref{fig6}c displays the image provided by the second singular vector $\mathbf{U_2}$. The latter image is much more {sparse} than the former one. {Notably}, $\mathbf{U_2}$ is often associated with the same coherent reflector as $\mathbf{U_1}$ but focuses on smaller details. Indeed, as soon as a scatterer is larger than a resolution cell, several singular vectors are associated with it. Their number $n$ actually scales as the number of resolution cells contained in the scatterer~\citep{aubry0,robert}. A weighted combination of singular vectors allows to recover the complete structure of the reflectors in the volcano:
\begin{equation}
\label{combine}
I(\mathbf{r})= \sum_{p=1}^n \lambda_p \left | U_p(\mathbf{r}) \right |^2 
\end{equation}
In the present case, only the two first singular vectors are considered ($n=2$) because the third singular value is dominated by the multiple scattering noise subspace of $\mathbf{R_F}$ (see \textit{e.g.} Figure~\ref{fig5}a). 
\begin{figure}[htbp]
 \centering
 \includegraphics[width=14cm]{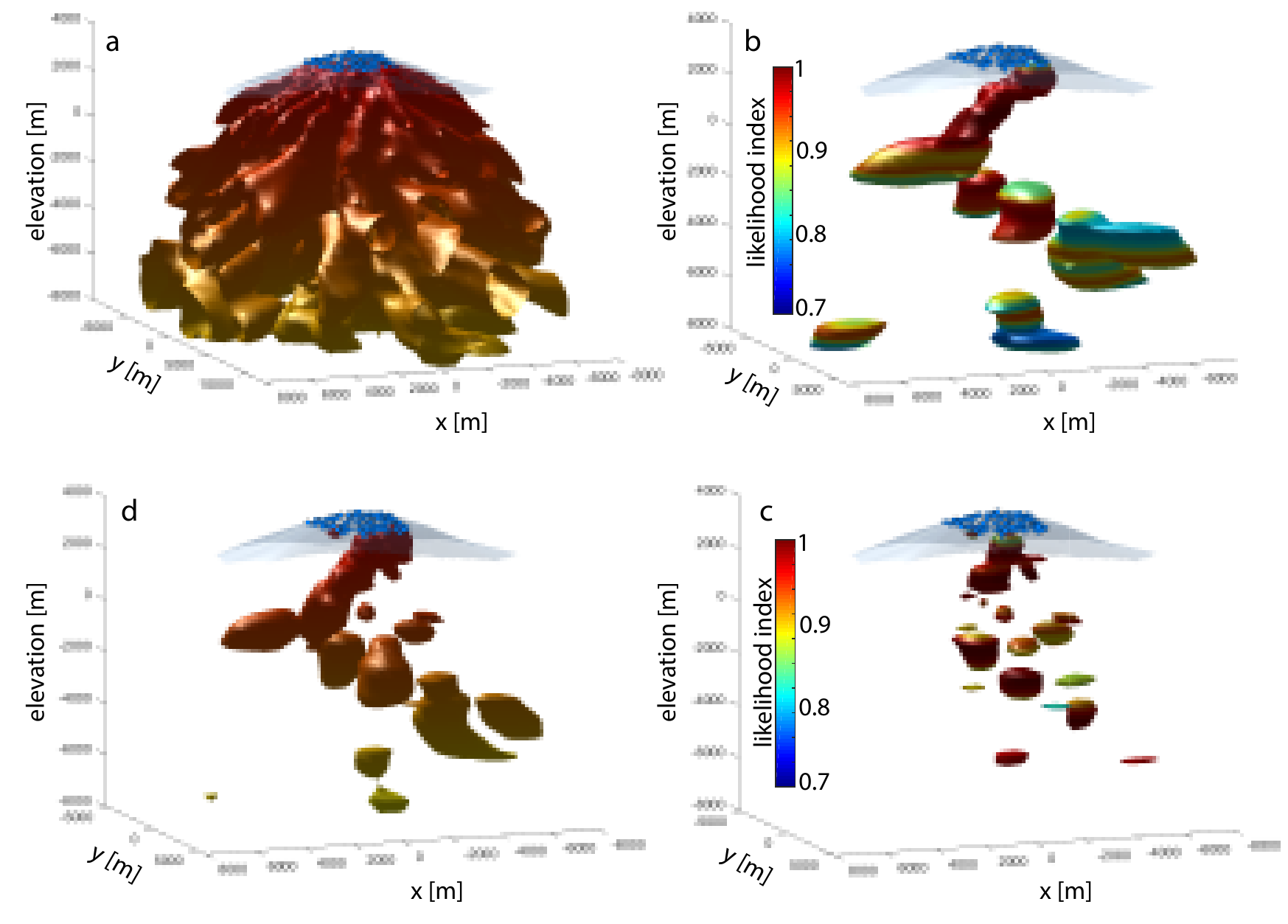}
 \caption{{Isosurface plots of the three dimensional images of the volcano built from (a) confocal focusing, (b,c) the first and second singular vectors of $\mathbf{R_F}$, $\lambda_1^2 |U_1(\mathbf{r})|^2$ and $\lambda_2^2|U_2(\mathbf{r})|^2$, respectively, and (d) the coherent combination of both [Eq.~(\ref{combine})]. For all these isosurface plots, the images are multiplied in depth by a factor $z^2$ to compensate for the geometrical decrease of singly-scattered body waves.The isosurface is fixed to be 4\% of the maximum of each image. The color scale in panels (b) and (c) accounts for the {likelihood index} associated with the singular vector at each depth (see Sec.\ref{rely}).}}
 \label{fig6}
  \end{figure}
	
Figure \ref{fig6}d displays the resulting three-dimensional image of the Erebus volcano. It should be compared with the random speckle image obtained by the direct confocal method (Figure~\ref{fig6}a).} The matrix approach described in this paper enables to highlight a chimney-shaped structure with a nearly 100\% {likelihood index} (Figure~\ref{fig6}b). Moreover, although computed independently, the results at each depth are spatially consistent, substantiating their physical meaning. \cor{In agreement with previous studies~\citep{Chaput2012,Zandomeneghi2013}, the chimney seems to rise towards the lava lake at shallow depth, followed by a sharp bifurcation towards the north-west, finally setting centrally in a shallow magma chamber near 2500 m elevation. Some particular features also emerge from the sea level to 5000 m below it. An interpretation of these imaged structures will be provided in the next section.}

Beforehand, however, it should be noted that the velocity model used for emission and reception is a {bulk approximation} and thus the depth is likely to be inaccurate. Yet a correct model would dilate the structure up or down but would not change significantly its geometrical shape. We should also keep in mind that the obtained 3D image still suffers from aberrations owing to the imperfect wave velocity model used for the Erebus volcano. One great advantage of iterative time reversal is the ability to detect scatterers in presence of the strong wave-front distortions. \cora{However, the associated image still suffers from aberrations as the wave velocity model used for the back-propagation is inaccurate. The perspective of this work will be to tackle these aberration issues. A first step will be to incorporate the existing wave velocity model over the first km depths~\citep{Zandomeneghi2013}, before improving and extending it to larger depths thanks to the reflection matrix approach. The removal of multiple reflections due to geological layering will be also a challenge~\citep{Verschuur}.}

\section{\label{interp} Structural interpretation}

\cor{A structural interpretation of the three-dimensional image displayed in Figure \ref{fig6} is now proposed by relying on the existing literature about Erebus. The Erebus magmatic system consists of a persistently open convecting lava lake, and has historically produced multi-phase magma, potentially a testament to its complex inner structure. Magmas related to the proposed hot spot upwelling beneath Ross Island have been found in wide regions (such as the Dry Valleys), and variability in the compositions have pointed to radial fracturing in the lower crust as a mechanism for diversification. All current models suggest a single original mantle source with subsequent crustal assimilation. A large mantle anomaly down to roughly 400 km depth has been imaged as a source~\citep{Gupta}. Beyond the modified confocal image produced by Ref.~\cite{Chaput2012} and the subsequent shallow active source tomography model~\citep{Zandomeneghi2013} however, there are no reliable structural images of Erebus at intermediate depths. This lack of information between 1-100 km is due to several factors.}

\cor{Firstly, there is a general lack of magmatic seismicity at Erebus, thus precluding event location studies, compounded with a paucity of tectonic earthquakes in West Antarctica that could be used in passive tomographic studies~\citep{Winberry}. Furthermore, given the extreme topographic expression of Erebus on Ross Island, ambient noise Rayleigh wave reconstructions for imaging at depths of interest would require stations located in the Ross Sea. Geochemical and geological methods therefore are the only alternative source of constraint for this depth range. Ref.~\cite{Moussalam}, for instance, studied phase equilibria in lavas for Erebus Lineage magmas (CO$_2$-H$_2$O-Pressure-Temperature-O$_2$-fugacity). These magma phase equilibria suggest that a pressurized deeper body capable of reproducing Erebus lavas must be contained in the 4-7.5 km range. Note that this is in agreement with the inferred deeper structures imaged in the 5 km depth range in Figure \ref{fig6}d.}

\cor{As a measure of confidence in our results, the top kilometer imaged in this study shows a distinctly West/NorthWest bifurcation immediately under the lava lake, which is supported by previous imaging studies~\citep{Chaput2012,Zandomeneghi2013}, as well as moment tensor inversions of Very Long Period (VLP) signals associated with gas slug movement in the shallow conduit preceding Strombolian eruptions~\citep{Aster}. The agreement between Ref.~\cite{Chaput2012} and this paper, both showing a West/NorthWest shallow bifurcation with a re-centralization at 2500 m elevation, is particularly encouraging, given that the former study was performed with autocorrelations of eruption coda, and thus an entirely different dataset. This corroboration lays credence to the unbiased nature of the icequake-reconstructed Green's functions.}

\cor{Furthermore, temporal variations in the timing of VLP signals with respect to the short period lava lake eruption imply concurrent changes in gas slug trapping mechanisms~\citep{Knox2018}, thus suggesting an abnormally lateral conduit system at shallow depths. This is a further testament to the complexity of the near surface conduit structure, and is reflected in the results presented here (see Figure\ref{fig6}d). Efforts are currently underway to back-project continuous seismic data from the ETS-ETB arrays~\citep{Hansen} and identify any potential coherent magmatic seismicity embedded in the noise that might locate at depth.}

	\section{Conclusion}
	
The analysis of the cross-correlation of coda waves recorded by a set of geophones under a reflection matrix approach provides an elegant and powerful tool for geophysical exploration. The reflection matrix actually contains all the information about the propagation of body waves beneath the geophones array. {The incoherent noise present in the cross-correlation of coda waves is efficiently reduced by the imaging procedure. This allows for an effective use of body waves for passive imaging with array correlations. Our approach first consists in applying a set of matrix operations }to perform an adaptive confocal selection of singly-scattered waves among a predominant multiple scattering background. Iterative time reversal is then applied to the focal reflection matrix in order selectively focus on the most scattering parts of the scattering medium while removing the residual multiple scattering noise. This approach is applied successfully to the case of Erebus volcano. A 3D image of the internal structure of the volcano is revealed while state-of-the art techniques fail in imaging such scattering media. Although the velocity model is approximate, the penetration depth and resolution are strongly enhanced compared to a direct back-propagation of seismic waves, and could be even improved using more sophisticated back-propagation tools.
Although only the single scattering contribution has been used here, a matrix analysis of the multiple scattering contribution can be fruitful to characterize the scattering and transport parameters of seismic waves propagating in the medium of interest~\citep{Aubry2007,Margerin2016,Mayor2017}. In any case, this approach is very general and can be particularly useful in strongly heterogeneous environments such as volcanoes, mountain areas or fault zones in which aberrations and scattering impede seismic bulk wave imaging. 

\section*{ACKNOWLEDGMENTS}
The authors are grateful for funding provided by LABEX WIFI (Laboratory of Excellence within the French Program Investments for the Future, ANR-10-LABX-24 and ANR-10-IDEX-0001-02 PSL*), the European Research Council (F-IMAGE) and by TOTAL R\&D. Portable seismic instruments for the TOMO Erebus experiment were provided by the Incorporated Research Institutions for Seismology (IRIS) through the PASSCAL Instrument Center at New Mexico Tech. Data are available through the IRIS Data Management Center under network code ZO (2011-2012), YA, and ZW (2007-2009). The facilities of the IRIS Consortium are supported by the National Science Foundation under cooperative agreement EAR-1063471, the NSF Office of Polar Programs, and the DOE National Nuclear Security Administration. 

\renewcommand{\thefigure}{S\arabic{figure}}
\setcounter{figure}{0}

\newpage

 \begin{figure}
 \noindent \includegraphics[width=17cm]{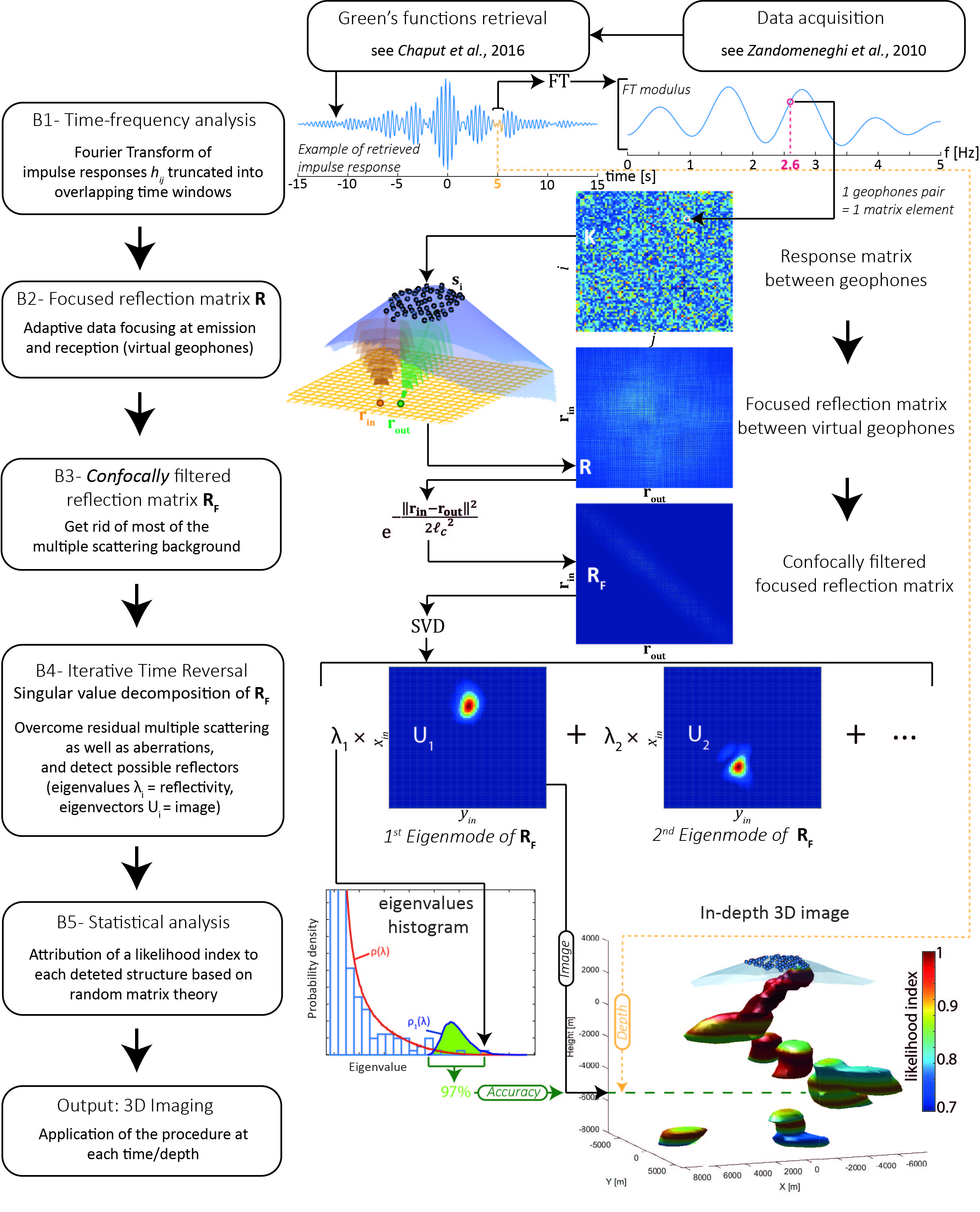}
\caption{Flow chart describing the matrix imaging process. (B0) The Green's functions between each geophone are recovered from coda based cross-correlation functions. (B1) The resulting impulse response matrix $\mathbf{H}$(t) is submitted to a time-frequency analysis, yielding the matrix $\mathbf{K}(t,f)$. (B2) Based on a rough estimate of velocity $c$, focusing is performed both at emission and reception at each time $t$ by means of simple matrix operations [Eq.~(4)]. It provides a new reflection matrix $\mathbf{R} $ defined for a set of virtual geophones located at depth $z=ct/2$. (B3) An input-output analysis, referred to as confocal filtering, allows for the removal of most of the multiple scattering contribution in the new reflection matrix $\mathbf{R_F}$ [Eq.~(6)]. (B4) Iterative time reversal is applied to overcome the residual multiple scattering contribution as well as the aberration effects induced by the scattering medium itself, and detect possible reflectors  [Eq.~(7)]. (B5) A statistical analysis of the matrix singular values permits attribution of a likelihood index to each detected structure [Eq.~(9)]. (B6) Significant eigenmodes of $\mathbf{R_F}$ obtained at each depth are stacked into a 3D image.}
 \label{figure_label}
 \end{figure}





\end{document}